# DenseGNN: universal and scalable deeper graph neural networks for high-performance property prediction in crystals and molecules

Hongwei Du, Jiamin Wang, Jian Hui*, Lanting Zhang*, Hong Wang*
1 School of Materials Science and Engineering, Shanghai Jiao Tong University, Shanghai 200240, China.
2 Zhangjiang Institute for Advanced Study, Shanghai Jiao Tong University, Shanghai 201203, China.
3 Materials Genome Initiative Center, Shanghai Jiao Tong University, Shanghai 200240, China.

Corresponding authors
Correspondence to: Hong Wang, hongwang2@sjtu.edu.cn (Hong Wang).

# Abstract

Modern generative models based on deep learning have made it possible to design millions of hypothetical materials. To screen these candidate materials and identify promising new materials, we need fast and accurate models to predict material properties. Graphical neural networks (GNNs) have become a current research focus due to their ability to directly act on the graphical representation of molecules and materials, enabling comprehensive capture of important information and showing excellent performance in predicting material properties. Nevertheless, GNNs still face several key problems in practical applications: First, although existing nested graph network strategies increase critical structural information such as bond angles, they significantly increase the number of trainable parameters in the model, resulting in a increase in training costs; Second, extending GNN models to broader domains such as molecules, crystalline materials, and catalysis, as well as adapting to small data sets, remains a challenge. Finally, the scalability of GNN models is limited by the over-smoothing problem. To address these issues, we propose the DenseGNN model, which combines Dense Connectivity Network (DCN), hierarchical node-edge-graph residual networks (HRN), and Local Structure Order Parameters Embedding (LOPE) strategies to create a universal, scalable and efficient GNN model. We have achieved state-of-the-art performance (SOAT) on several datasets, including JARVIS-DFT, Materials Project, QM9, Lipop, FreeSolv, ESOL, and OC22, demonstrating the generality and scalability of our approach. By merging DCN and LOPE strategies into GNN models in computing, crystal materials and molecules, we have improved the performance of models such as GIN, Schnet and Hamnet on materials datasets such as Matbench. The LOPE strategy optimizes the embedding representation of atoms and allows our model to train efficiently with a minimal level of edge connections. This substantially reduces computational costs and shortens the time required to train large GNNs while maintaining accuracy. Our technique not only supports building deeper GNNs and avoids performance penalties experienced by other models, but is also applicable to a variety of applications that require large deep learning models. Furthermore, our study demonstrates that by using structural embeddings from pre-trained models, our model not only outperforms other GNNs in distinguishing crystal structures but also approaches the standard X-ray diffraction (XRD) method.

# Introduction

In the almost infinite design space of chemistry, only $10^5$ crystal structures have been synthesized and characterized, forming a very limited region of the potential material space. To push the boundaries of existing material properties and explore a broader material design space, one of the most promising approaches is the generative design paradigm based on deep learning (DL). In this approach, existing materials are fed into deep generative models based on neural networks, which learn atomic assembly rules to form stable crystal structures and use these rules to generate chemically viable hypothetical structures or compositions[1-3]. Although these material candidates can be rapidly generated in the hundreds or thousands, a fast and accurate model for predicting material properties is required to screen the most promising materials for further description of their properties, whether through first-principles density functional theory (DFT) or molecular dynamics (MD) calculations or through experiments. After all these steps, unique materials can be found in the unknown design space.

Currently, machine learning (ML) models have become one of the most promising methods in materials discovery, offering higher prediction accuracy and speed compared to first-principles calculations[4]. ML models based on composition or structure can successfully predict material properties, with their performance heavily influenced by the choice of ML algorithms, features, and the quality and quantity of available datasets. Among these screening models, composition-based ML models[5-7] have the advantage of speed and the ability to screen large-scale hypothetical compositions generated by DL models[8]. However, almost all material properties strongly depend on the structure of the material, so structure-based material prediction models typically have higher prediction accuracy. They can be used to screen known material structure repositories, such as ICSD[9] or the (MP) Database[10], or the structures of hypothetical crystal materials created by modern generative DL models[11-14]. Currently, there are two main classes of ML methods for predicting material properties based on structure, which are divided into those based on heuristic features and those based on DL models that learn features. Although heuristic feature-based ML models[15,16] have shown some success in various applications, such as formation energy prediction[17] and ion conductivity screening[18], extensive benchmark studies have shown that GNNs outperform them in material performance prediction[19]. GNNs are used to process graph-structured data and are closely related to geometric deep learning. In addition to research on social and citation networks and knowledge graphs, chemistry is one of the main driving forces behind GNN development. GNNs can be viewed as an extension of convolutional neural networks to handle irregularly shaped graph structures. Their architecture allows them to directly work on natural input representations of molecules and materials, which are chemical graphs composed of atoms and bonds, or even the three-dimensional structure or point cloud of atoms. Therefore, GNNs can fully represent materials at the atomic level[20] and have the flexibility to incorporate physical laws[21] and phenomena at larger scales, such as doping and disorder. Using this information, GNNs can learn valuable and information-rich internal material representations for specific tasks, such as predicting material properties. Thus, GNNs can complement or even replace handcrafted feature representations widely used in the natural sciences.

Since 2018, various GNN models have been proposed to improve prediction performance, such as CGCNN[22], SchNet[20], MEGNet[23], iCGCNN[24], ALIGNN[25], and coGN[26], among others.

These architectures all use structure graph representation as input, also incorporating slightly different additional information, convolution operators, and neural network architectures[20,23,24]. Despite this progress, there are still major problems with the application of GNNs in the fields of chemistry and materials: First, nested graph networks like ALIGNN and coNGN[26] have substantially more trainable parameters, leading to higher training costs compared to non-nested graph networks. These nested graph networks maintain their advantages only on some crystal datasets. Therefore, it is necessary to develop strategies that can effectively embed information on many-body interactions such as bond angles and local geometric distortions outside of nested graph networks. Second, there exists an imbalance in research and model development efforts across the fields of materials science, molecular science, and chemistry. Extending existing GNN models to broader application areas (such as spanning molecular, crystal materials, and catalysis fields) may be challenging and require further development of GNN models. At the same time, the data released by Matbench official website[27] currently shows that GNNs generally perform worse on small datasets compared to the MODNet[28], which incorporates traditional feature engineering[29]. Thirdly, the strategy for constructing graph representations of input structures is a key factor affecting the performance and training cost of GNNs. The coGN proposes asymmetric unit cells as representations, reducing the number of atoms by utilizing all symmetries of the crystal system to minimize the number of nodes in the crystal graph. This reduces the time required to train large GNNs without sacrificing accuracy. However, the method of reducing the number of atoms does not effectively optimize edge connections in some material datasets, such as the perovskites dataset in Matbench, where using all symmetry does not reduce the average number of nodes. Finally, most message-passing GNNs currently suffer from oversmoothing problems, in which the representation vectors of all nodes of a graph become indistinguishable as the number of graph convolution (GC) layers increases[30-33], limiting the increase in GC layers. As the number of layers increases, the model performance decreases.

In this work, we propose DenseGNN, a GNN model that combines DCN, HRN, and LOPE to overcome oversmoothing problems, support the construction of very deep GNNs, and avoid the performance degradation problems present in other models. Our model allows for highly efficient training at the level of minimal edge connections. We also apply the DCN and LOPE strategies to GNNs in the fields of computers, crystal materials, and molecules, achieving performance improvements on almost all models in the Matbench dataset. Our contributions in this paper can be summarized as follows:

- We overcome the main bottlenecks of GNNs in predicting material properties and propose a novel DCN-based GNN architecture. This architecture updates edge, node, and graph-level features simultaneously during the message-passing process through DCN and residual connection strategies, achieving more direct and dense information propagation, reducing information loss during propagation in the network, overcoming oversmoothing problems, and supporting the construction of very deep GNNs. Additionally, it better utilizes feature representations from preceding layers, improving network performance and generalization.
- DenseGNN outperforms the latest coGN, ALIGNN, and M3GNet[34] on most benchmark datasets for crystal, and molecules materials, while also demonstrating higher learning efficiency on experimental small datasets.
- We apply our DCN and LOPE strategies to GNNs in fields such as computers, crystal materials, and molecules, achieving notable performance improvements on all GNNs in the Matbench material dataset.
- Many important material properties (especially electronic properties such as band gaps) are very sensitive to structural features such as bond angles and local geometric distortions. Therefore, effectively learning these many-body interactions is crucial. Current strategies mainly involve building nested graph networks based on bond graphs to introduce bond angle information, but this method has high training costs. By incorporating LOPE and optimizing atomic embeddings, we minimize edge connections, reduce training time for large GNNs while maintaining accuracy.
- We demonstrate the improvement in the ability to distinguishing crystal structures by utilizing pre-trained model structural embeddings compared to other GNNs, approaching the standard XRD method.

# Results

## Model architecture description

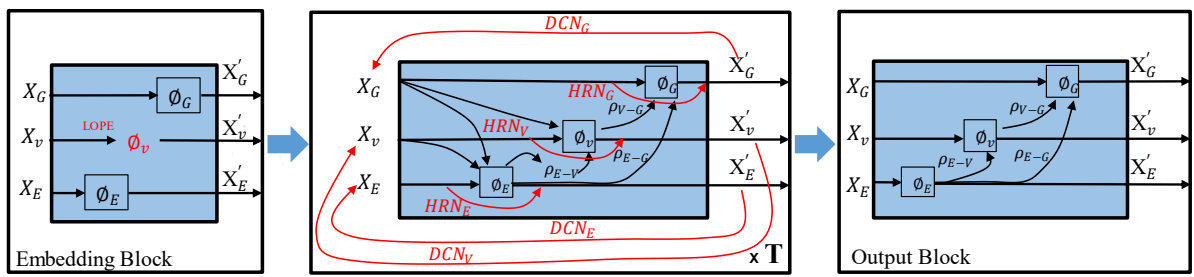

Figure1, An overview of the DenseGNN architecture.

In the architecture of DenseGNN, the first block is an edge-node-graph input embedding block, as shown in Figure1, which independently embeds atom/node, bond/edge, and global state/graph features. We employed k-nearest neighbors (KNN) as a default preprocessing method for edge selection, with the parameter k set to either 12 or 6 to achieve optimal test results. Figure 1 illustrates the high-level architecture of the GNN, but does not specifically instantiate the update functions $\phi_E$, $\phi_V$, $\phi_G$, and aggregation functions $\rho_{E \to V}$, $\rho_{V \to G}$, $\rho_{E \to G}$. The $\phi_E$ function utilizes 32 Gaussian functions uniformly distributed in the [0, 8] Å range to expand edge distances. As only distance information is used, this embedding is E(3)-invariant. The initial representations are projected into a 128-dimensional embedding space and implemented through a single Multilayer Perceptron (MLP) network. The $\phi_V$ function embeds atomic features (including LOPE, atomic number, atomic mass, atomic radius, ionization state, and oxidation state) into a 128-dimensional space. The $\phi_G$ function updates the attributes at the molecule/crystal level or state (e.g., the system's temperature). A more detailed explanation of the embedding block can be found in the Figure2.

In the second block of the Figure1, we implemented a sequentially connected structure consisting of T=5 GC processing blocks, each with independent learnable parameters and identical configuration. In each block, edge, node, and graph-level features are updated synchronously and are connected through DCNs respectively, achieving a comprehensive optimization of the local chemical environment of the atoms. A more detailed explanation of the edge-node-graph update can be found in the Figure3.

In the third block of the Figure1, we designed an independent readout module that aggregates node, edge, and graph features into graph-level features and inputs them into a single-layer MLP with a linear activation function to generate the final predictions of crystal properties. Throughout the hidden representations of the GNN, we uniformly used a 128-dimensional feature space. Unless specified otherwise, we used the common swish activation function in the MLPs. The GNN is trained using an Adam optimizer with a linear learning rate scheduler for 300 epochs.

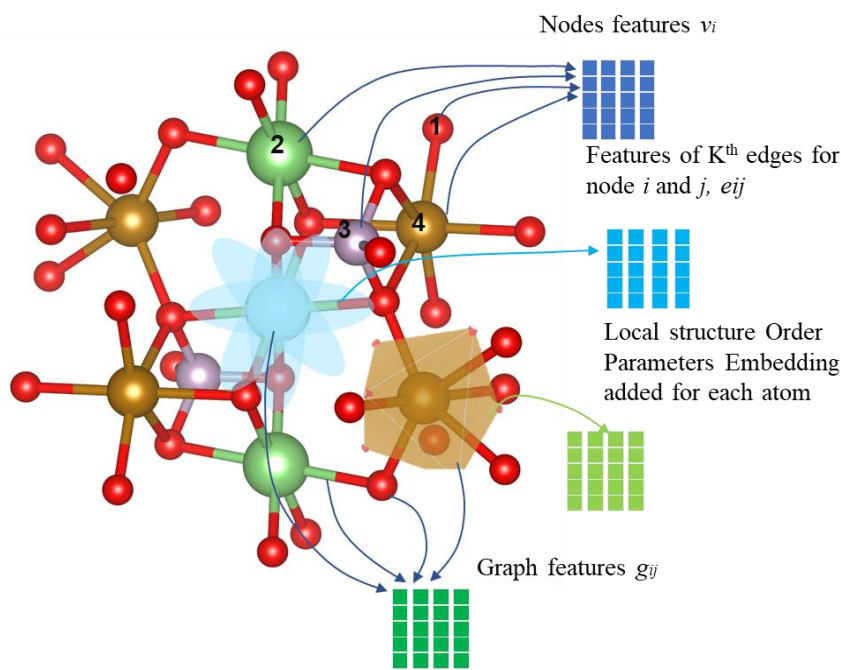

Figure 2, Representation of the local environment of atoms in a crystal structure. The nodes and edges are embedded with vectors that characterize the constituent atoms, the LOPE of atoms and their correlations with neighboring atoms. graph state vector storing the molecule/crystal level or state attributes.

Figure 2 elaborates on the embedding block part of Figure 1, illustrating the representation of the local environment around atoms in a crystal structure. In this crystal graph, nodes and edges are embedded with vectors, which characterize the constituent atoms and their correlations with neighboring atoms. The local chemical environment of the nodes is represented by concatenating the features of the constituent atoms and the LOPE. The edge vectors capture the local structural information of the crystal graph by selecting k-nearest-neighbors or using a parameter-free Voronoi method. Each edge is also embedded with a vector $e^k(i,j)$, which contains the distance information between adjacent atoms $i$ and $j$ within the crystal unit cell. To account for the periodicity of the crystal, multiple edges between atoms $i$ and $j$ can exist, indexed by $k$. Each node in the crystal graph is connected to its 6 or 12 nearest neighbors (DenseGNN performs better under this parameter.). Finally, $g$ is a graph state vector storing the molecule/crystal level or state attributes (e.g.,the temperature, charge of the system). In the input part of DenseGNN, the graph state vector is not mandatory and can be omitted depending on the specific use case.

The LOPE feature reflects the local environment and coordination of atoms around specific positions in the materials/molecular system. It consists of atomic embeddings and orientation-resolved embeddings, obtained by calculating the product integral of the radial distribution function (RDF) and a Gaussian window function. The atomic embeddings provide information about the local atomic environment, such as the density and distribution of neighboring atoms, by weighted summing the distances between the central atom and adjacent atoms. The orientation-resolved embeddings take into account the direction of neighboring atoms in relation to the central atom, providing a more detailed description of the local atomic environment, which includes orientation and anisotropy information. By introducing LOPE, we optimized the embedding representation of atoms, which is different from the nested graph strategy of ALIGNN as it includes bond angle information. This method avoids the high training cost of nested graph network strategies like ALIGNN, DimeNetPP and coNGN while maintaining model accuracy, thereby improving training efficiency.

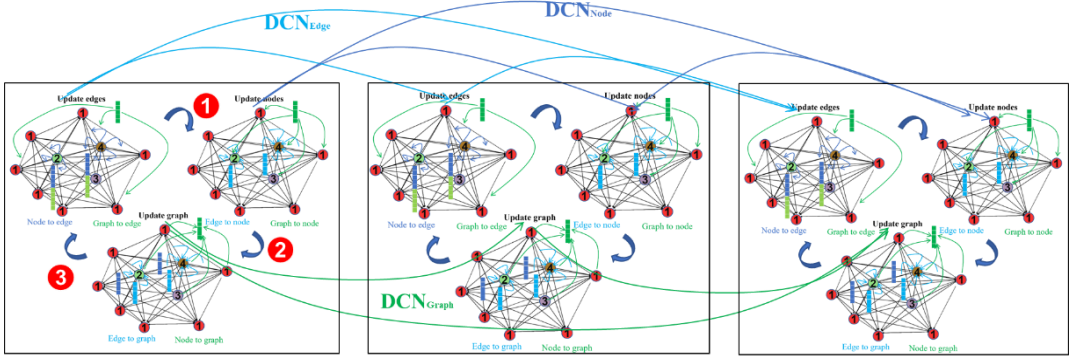

Figure 3, The input graph for DenseGNN consists of node attributes, edge attributes, and graph attributes (the graph attribute is not mandatory). In the first update step, edge attributes are updated based on information from the nodes forming the edges, the graph attributes, and the previous edge attributes. Subsequently, the second and third steps update the node and graph attributes, respectively, through information flow among all three attributes.

$$e_{ij}^{t+1} = e_{ij}^t + \sum_{j,k} \sigma([v_i^t \oplus v_j^t \oplus u_{ij}^t \oplus e_{ij}^t]W_1^t + b_1^t) \quad (1)$$

$$v_i^{t+1} = v_i^t + \sum_{j,k} \sigma([v_i^t \oplus u_{ij}^t \oplus e_{ij}^t]W_2^t + b_2^t) \quad (2)$$

$$u_{ij}^{t+1} = u_{ij}^t + \sum_{j,k} \sigma([v_i^t \oplus u_{ij}^t \oplus e_{ij}^t]W_3^t + b_3^t) \quad (3)$$

Equations 1, 2, and 3 correspond to $\phi_E$, $\phi_V$ and $\phi_G$ in Figure1, respectively, detailing the edge-node-graph update process in DenseGNN during training. Below, we will explain the update process step by step.

1. Edge feature update: The edge update function $\phi_E$ combines features of the receiver and sender nodes, global state, and edges, transforms them through a three-layer MLP network, and updates edge features through residual networks and activation functions.
2. Node feature update: For each node, we first aggregate the updated edge features, then concatenate the aggregated node features with the global state and original node features, transform them through a single-layer MLP network $\phi_V$, and update node features through residual networks and activation functions.
3. Graph feature update: For each graph, we sum or average the updated edge and node features, then concatenate the aggregated graph features with the global state, transform them through a single-layer MLP network $\phi_G$, and update graph features through residual networks and activation functions. To ensure efficient flow of information between blocks, we adopted the design concept of DCN, directly connecting all blocks to each other.

Each block not only receives features from all preceding blocks but also passes its own node, edge, and graph features to subsequent blocks. This design introduces *L(L+1)/2* connections in the network, rather than the traditional *L* connections in the architecture. This feature of DCN not only improves the flow of information and gradients, making the network easier to train, but also allows each layer to directly access gradients from the loss function and original input signal, thereby achieving implicit deep supervision[35]. This helps in training deeper network architectures and overcoming over-smoothing problems in GNNs. Additionally, dense connections have a regularization effect, helping to reduce overfitting on tasks with small training set sizes. Figures 1 and 12 schematically illustrate this architectural layout.

We noticed that dense connections may slow down model inference speed. However, factors affecting model training and inference speed primarily include three aspects: the method for constructing crystal/molecular graphs, the model's hyperparameter settings (such as learning rate and batch size), and the model's training parameters. On the Matbench and Jarvis DFT datasets, we compared DenseGNN with reference models and found that, with consistent hyperparameter settings, regardless of whether the radius-based or k-nearest-neighbors method is chosen, DenseGNN requires fewer edges than reference models such as MEGNet, SchNet, CGCNN, and coGN. This compensates for the dense connections in the DCN. Additionally, in Supplementary Figure S1, we provide a low-parameter version of DenseGNN, DenseGNN-Lite, maintaining the DCN and LOPE strategies, using a crystal graph based on KNN, and no longer using the graph state. By optimizing the edge-node update strategy, we significantly reduce the model's parameters while only slightly decreasing the model's performance, which remains superior to recent coGN in most cases.

Overall, the DenseGNN stands out for its simplicity, containing only MLP as update functions and mean or max aggregation functions, without the complex edge gate control or attention-based message passing mechanisms found in CGCNN, ALIGNN, or GeoCGNN[36]. By introducing the mechanism of DCN, we successfully constructed a very deep GNN architecture while avoiding performance degradation problems, enhancing the scalability of GNNs.

# Model comparison and analysis

| Property | Dense GNN | Dense GNN-Voronoi | Dense GNN-Lite | DenseGNN-Voronoi-Lite | Dense NGN | ALIGNN | Schnet (kgcnn) | M3GNet | MODNet | coGN | coNGN | #Crystals | %Improve |
|---|---|---|---|---|---|---|---|---|---|---|---|---|---|
| Train parameters | 1717889 | 1721089 | 771456 | 775808 | 3101253 | 7355236 | 316225 | * | * | 697601 | 2799201 | * | * |
| Jdft2d | 32.5425±7.039 | 31.8491±9.1139 | 32.7446±9.3890 | 32.2381±8.996 | **30.2201±8.0520** | 43.4240±8.949 | 42.6637±13.720 | 50.1719±11.90 | 33.192±7.343 | 37.1652±13.683 | 36.170±11.597 | 636 | 8.95 |
| Phonons | 24.8470±1.778 | 24.6663±2.2923 | 24.9106±1.3046 | 24.9987±1.496 | **23.5230±1.3923** | 29.5390±2.115 | 38.9636±1.9760 | 34.1606±4.500 | 34.2751±2.078 | 29.7117±1.997 | 28.887±3.284 | 1265 | 18.57 |
| Dielectric | 0.2837±0.0633 | 0.2901±0.0420 | 0.2967±0.0544 | 0.2887±0.0464 | 0.287±0.0452 | 0.3449±0.0871 | 0.3277±0.0829 | 0.3120±0.063 | **0.2711±0.0714** | 0.3088±0.0859 | 0.3142±0.074 | 4764 | -4.64 |
| Perovskites | **0.0268±0.0007** | 0.0271±0.0011 | 0.0286±0.0011 | 0.0274±0.0007 | 0.0271±0.0008 | 0.0288±0.0005 | 0.0342±0.0001 | 0.0330±0.0028 | 0.0908±0.0028 | 0.0269±0.0009 | 0.0290±0.0011 | 18,928 | 0.37 |
| Log gvrh | 0.0668±0.0012 | 0.0660±0.0014 | 0.0724±0.0007 | 0.0687±0.0011 | **0.0654±0.0009** | 0.0715±0.0006 | 0.0796±0.0022 | 0.0860±0.002 | 0.0731±0.0007 | 0.0689±0.0009 | 0.0670±0.0006 | 10,987 | 2.39 |
| Log kvrh | 0.0512±0.0009 | 0.0509±0.0012 | 0.0518±0.0010 | 0.0505±0.0013 | **0.0464±0.0019** | 0.0568±0.0028 | 0.0590±0.0022 | 0.0580±0.0028 | 0.0548±0.0025 | 0.0535±0.0028 | 0.0491±0.0026 | 10,987 | 5.50 |
| Band gap | 0.1551±0.0031 | **0.1548±0.0023** | 0.1721±0.0032 | 0.1617±0.0026 | 0.1590±0.0024 | 0.1861±0.0030 | 0.2352±0.0034 | 0.1830±0.0050 | 0.2199±0.0059 | 0.1559±0.0017 | 0.1697±0.0035 | 106,113 | 0.71 |
| Formation enthapy | 0.0179±0.0002 | **0.0166±0.0002** | 0.0185±0.0004 | 0.0180±0.0002 | 0.0170±0.0006 | 0.0215±0.0005 | 0.0218±0.0004 | 0.01950±0.0002 | 0.0448±0.0039 | 0.0170±0.0003 | 0.01780±0.0004 | 132,752 | 2.35 |

In Figure 4, The MAE results of different versions of DenseGNN on the MatBench datasets are compared with those of previous models including SchNet, MODNet and ALIGNN, as well as recent models such as M3GNet, coGN, and coNGN. Unlike other baseline models, ALIGNN, coNGN, and M3GNet belong to the category of nested graph networks and incorporate angle information. The properties evaluated include e_form (eV/atom), gap (eV), perovskites (eV/unit cell), log_kvrh (log10(GPa), log_gvrh (log10(GPa)), dielectric (unitless), phonons (1/cm), and jdft2d (meV/atom). The best results, data size and relative improvement are highlighted in bold. * denotes that the training parameters were not provided or no training parameters.

| QM9/Models | units | Dense GNN | Dense NGN | MEGNet | SchNet | enn-s2s | ALIGNN | DN++ |
|---|---|---|---|---|---|---|---|---|
| $\epsilon_{HOMO}$ | eV | 0.0261 | **0.0211** | 0.043 | 0.041 | 0.043 | 0.0214 | 0.0246 |
| $\epsilon_{LUMO}$ | eV | 0.0216 | **0.0181** | 0.044 | 0.034 | 0.037 | 0.0195 | 0.0195 |
| $\Delta\epsilon$ | eV | 0.0423 | **0.0305** | 0.066 | 0.063 | 0.069 | 0.0381 | 0.0326 |
| ZPVE | meV | 1.35 | 1.24 | 1.43 | 1.70 | 1.50 | 3.10 | **1.21** |
| $\mu$ | D | 0.0422 | 0.0234 | 0.050 | 0.033 | 0.030 | **0.0146** | 0.0297 |
| $\alpha$ | bohr$^3$ | 0.0539 | 0.0475 | 0.081 | 0.235 | 0.092 | 0.0561 | **0.0435** |
| $\langle R^2 \rangle$ | bohr$^2$ | 0.1404 | 0.1056 | 0.302 | **0.073** | 0.180 | 0.5432 | 0.331 |
| $U_0$ | eV | 0.00828 | **0.00584** | 0.012 | 0.014 | 0.019 | 0.0153 | 0.00632 |
| U | eV | 0.00810 | 0.00650 | 0.013 | 0.019 | 0.019 | 0.0144 | **0.00628** |
| H | eV | 0.00846 | **0.00599** | 0.012 | 0.014 | 0.017 | 0.0147 | 0.00653 |
| G | eV | 0.00819 | **0.00697** | 0.012 | 0.014 | 0.019 | 0.0144 | 0.00756 |
| $C_v$ | cal (mol K)$^{-1}$ | **0.0225** | 0.0237 | 0.029 | 0.033 | 0.040 | - | 0.0230 |

Figure 5, Test MAE for multiple tasks of the QM9 datasets. Models for comparison are DenseGNN, DenseNGN, MEGNet, SchNet, enn-s2s, ALIGNN and DimeNet++ (DN++). This includes DenseNGN, which is a nested graph network that incorporates angle information, enabling the model to learn more accurate local chemical environment information. The best results are indicated in bold font. DenseNGN implements the DCN strategy within the nested graph networks framework of coNGN. - denotes that the MAE results were not provided.

| Property | %Improve | DenseGNN | DenseGNN-Voronoi | DenseLite | DenseLite Voronoi | coGN | CGCNN | Schnet kgcnn | Megnet kgcnn | Matminer | CFID | Train/val/test |
|---|---|---|---|---|---|---|---|---|---|---|---|---|
| Train parameters | * | 1717889 | 1721089 | 771456 | 775808 | 697601 | 369984 | 316225 | 155073 | * | * | * |
| dft_3d_mepsz | 8.11 | 22.7585 | 22.4544 | 22.4477 | **22.1535** | 24.1081 | 36.0538 | 25.668 | 27.292 | 24.6651 | 29.3445 | 13447/1681/1681 |
| dft_3d_exfoliation_energy | 34.91 | 34.2668 | 32.1132 | 34.4445 | **31.0476** | 47.6979 | 52.7033 | 48.3027 | 68.2435 | 40.887 | 62.1169 | 650/81/81/812 |
| dft_3d_shear_modulus_gv | 5.18 | 8.2219 | **8.2124** | 8.6448 | 8.441 | 8.6612 | 16.0459 | 10.2291 | 10.4359 | 10.5415 | 11.9164 | 15744/1968/1968 |
| dft_3d_spillage | 4.27 | **0.3455** | 0.3548 | 0.3541 | 0.3496 | 0.3609 | 0.3965 | 0.3622 | 0.3904 | 0.3592 | 0.3867 | 9101/1137/1137 |
| dft_3d_optb88vdw_total_energy | 4.20 | 0.0259 | **0.0251** | 0.0289 | 0.0291 | 0.0262 | 0.0815 | 0.0374 | 0.0393 | 0.0936 | 0.2436 | 44569/5572/5572 |
| dft_3d_mepsx | 4.67 | 23.9040 | 23.7881 | 23.7232 | **23.0981** | 24.2289 | 33.6597 | 26.0289 | 26.6785 | 25.2932 | 30.261 | 13447/1681/1681 |
| dft_3d_epsz | 6.62 | 18.5236 | 18.4360 | 18.6523 | **18.3210** | 19.6192 | 33.6597 | 21.5016 | 22.6781 | 25.2932 | 24.781 | 35592/4449/4449 |
| dft_3d_dfpt_piezo_max_dij | 3.79 | 16.0002 | **15.8002** | 16.1023 | 15.7891 | 15.2235 | 16.0135 | 18.6753 | 15.7584 | 21.5729 | - | 2677/334/334 |
| dft_3d_mepsy | 6.07 | 23.1297 | 23.0134 | 24.0287 | **22.722** | 24.1891 | 32.4577 | 25.5455 | 22.6781 | 25.0706 | 30.0578 | 13447/1681/1681 |
| dft_3d_kpoint_length_unit | -2.04 | 9.3047 | 9.2319 | 9.5510 | 9.2647 | 9.5722 | 13.2145 | 10.102 | 10.3826 | **9.0470** | 9.7085 | 44313/5540/5540 |
| dft_3d_n_powerfact | 6.42 | 441.5224 | 435.5561 | 431.1588 | **423.2055** | 452.235 | - | 495.4136 | 501.3722 | 469.6279 | - | 18568/2321/2321 |
| dft_3d_ph_heat_capacity | 13.40 | 5.4538 | **5.2933** | 6.8319 | 6.6523 | 6.1125 | - | 17.7707 | 6.0443 | 5.2757 | - | 9644/1205/1205 |
| dft_3d_formation_energy_peratom | 7.38 | 0.0269 | **0.0258** | 0.0276 | 0.0270 | 0.0271 | 0.0625 | 0.0365 | 0.0423 | 0.0734 | 0.1419 | 44569/5572/5572 |
| dft_3d_epsx | 4.60 | 19.6571 | **19.0801** | 19.810 | 19.4701 | 20.0004 | 31.4744 | 22.0798 | 21.775 | 21.2597 | 24.8408 | 35592/4449/4449 |
| dft_3d_optb88vdw_bandgap | 0.90 | 0.1210 | 0.1212 | 0.1215 | **0.1208** | 0.1219 | 0.1908 | 0.698 | 0.146 | 0.1873 | 0.299 | 44569/5572/5572 |
| dft_3d_max_efg | 13.23 | 18.7130 | 18.4835 | 18.7936 | **17.7365** | 20.4417 | 24.6695 | 23.4912 | 23.0652 | 19.4382 | - | 9493/1186/1186 |
| dft_3d_epsy | 21.81 | 19.6146 | **18.9137** | 19.6062 | 19.3089 | 24.1891 | 32.4577 | 22.1843 | 22.4066 | 25.0706 | 30.0578 | 35592/4449/4449 |
| dft_3d_encut | 14.24 | 129.0896 | **114.8239** | 122.0318 | 116.0048 | 133.8915 | 190.3857 | 253.3669 | 139.6071 | 138.2769 | 139.4357 | 44308/5539/5539 |
| dft_3d_n_Seebeck | -0.76 | 39.8638 | 39.5683 | 41.1822 | 39.8599 | **39.2692** | 49.3172 | 47.244 | 47.2813 | 44.2229 | - | 18568/2321/2321 |
| dft_3d_ehull | 9.66 | 0.0428 | **0.0421** | 0.0589 | 0.0512 | 0.0466 | 0.173 | 0.1014 | 0.0569 | 0.0601 | - | 44290/5537/5537 |
| dft_3d_bulk_modulus_kv | 1.85 | 8.9526 | **8.8260** | 8.9953 | 8.9650 | 8.992 | 19.3028 | 10.7105 | 11.4287 | 12.7411 | 14.1999 | 15744/1968/1968 |
| dft_3d_avg_hole_mass | 10.28 | 0.1289 | **0.1231** | 0.1308 | 0.1233 | 0.1372 | - | 0.1459 | 0.1416 | 0.1529 | - | 14114/1764/1764 |
| dft_3d_avg_elec_mass | 11.99 | 0.0832 | 0.0830 | 0.0830 | **0.0807** | 0.0917 | - | 0.0866 | 0.0896 | 0.107 | - | 14114/1764/1764 |
| dft_3d_mbj_bandgap | 1.59 | **0.2598** | 0.2829 | 0.2745 | 0.2662 | 0.264 | 0.4067 | 0.3289 | 0.297 | 0.3392 | 0.5313 | 14535/1817/1817 |
| dft_3d_dfpt_piezo_max_dielectric | -6.08 | 28.2282 | 28.9025 | 28.7372 | 28.1397 | 30.2923 | 32.5589 | **26.5276** | 30.1911 | 36.6913 | - | 3764/470/470 |
| dft_3d_slme | 2.78 | 4.4345 | 4.4621 | 4.5030 | **4.3268** | 4.4507 | 5.6603 | 5.2322 | 5.0614 | 4.9255 | 6.2607 | 7250/906/906 |
| dft_3d_magmom_oszicar | 2.92 | 0.2487 | **0.2429** | 0.2747 | 0.2611 | 0.2502 | 0.3543 | 0.7755 | 0.7753 | 0.3645 | 0.4748 | 41766/5222/5222 |
| qe_tb_f_enp | 7.01 | 0.0934 | 0.0901 | 0.0912 | **0.0889** | 0.0956 | - | 1.1637 | - | 0.3219 | - | 663659/82957/82957 |
| qe_tb_final_energy | 2.94 | 1.2988 | **1.2798** | 1.2957 | 1.2891 | 1.3185 | - | 86.2102 | - | 1.4714 | - | 663659/82957/82957 |
| qe_tb_indir_gap | 15.40 | 0.0412 | **0.0401** | 0.0465 | 0.0422 | 0.0474 | - | 0.11 | - | 0.0351 | - | 663659/82957/82957 |
| qe_tb_energy_per_atom | 5.03 | 0.0620 | **0.0604** | 0.0632 | 0.0625 | 0.0636 | - | 16.3509 | - | 1.5049 | - | 663659/82957/82957 |

Figure 6, Test MAE for multiple tasks of the JARVIS-DFT datasets. The MAE results of different versions of DenseGNN are compared with those of previous models including coGN, CGCNN, Matminer and CFID. These models do not belong to the category of nested graph networks and do not include angle information. The best results and relative improvement are highlighted in bold. - denotes training failure due to insufficient computational resources. * denotes that the training parameters were not provided or no training parameters. The figure provides the train/val/test split ratios for each dataset.

To ensure the performance evaluation of DenseGNN is both fair and accurate, we have taken a series of measures to ensure comparisons with other SOTA models are conducted under equal conditions. This includes training all models on the same dataset, using appropriate training and optimal hyperparameters, and employing the same cross-validation methods. We implemented all models based on the Keras Graph Convolution Neural Networks (KGCNN)[37] framework and set the hyperparameters for all models in a unified json configuration file. All models' hyperparameters are specified in JSON files, which can be referenced at https://github.com/dhw059/DenseGNN/tree/main/training/hyper. All models were trained for 300 epochs. To comprehensively evaluate the performance of different versions of DenseGNN, we tested them on multiple datasets, including molecular property datasets (QM9, LipopDataset, FreeSolvDataset, and ESOLDataset)[38-42], catalysis datasets (Open Catalyst Project, OC22)[43], and solid-state material datasets (Matbench and JARVIS-DFT)[27,44]. These datasets are widely used to evaluate the performance of models in various material property prediction tasks, with detailed descriptions provided in the dataset description section. For the Matbench dataset, we adopted the official default training-validation-testing split strategy. As for the JARVIS-DFT and QM9 datasets, we employed an 80%:10%:10% split strategy, aligning with the split used for the coGN and MEGNet datasets. The difference between DenseGNN-KNN and DenseGNN-Voronoi lies in the edge selection method, with the former using a k-nearest-neighbors approach and the latter a Voronoi-based approach. DenseGNN-Lite, as shown in Figure S1, is a streamlined version of DenseGNN that maintains performance while substantially reducing the number of training parameters.

The comparison results on the 8 regression task datasets in Matbench are shown in Figure 4, using the mean absolute error (MAE) metric. The results for ALIGNN, SchNet, M3GNet, MODNet, coGN, and coNGN are from previous benchmark studies. DenseGNN and DenseGNN-Voronoi performed better on almost all material property prediction tasks, especially showing substantial advantages on small datasets like Jdft2d, Phonons, and Dielectric. It is worth noting that, compared to ALIGNN, MODNet, and coNGN, the architecture of DenseGNN neither introduces bond angles through nested graph networks like ALIGNN and coNGN, nor incorporates traditional domain knowledge databases like Matminer[29] at input as MODNet does. Instead, it cleverly fuses DCN, HRN, and the LOPE strategy, optimizing the network structure for efficient information flow and feature reuse. By introducing atomic embeddings for local chemical environment information, DenseGNN avoids increasing training costs, enhances model performance.

Evaluation on the QM9 molecular property dataset (130,829 molecules) showed that DenseGNN achieved competitive results on multiple tasks compared to other reference models, such as SchNet, MEGNet, enn-s2s[45], ALIGNN, and DimeNet++[46], especially in tasks like Highest Occupied Molecular Orbital (HOMO) and energy gap ($\Delta\epsilon$), as shown in Figure 5. DenseGNN outperformed ALIGNN in most tasks and approached the performance of DimeNet++. DenseNGN implements the DCN strategy within the nested graph networks framework of coNGN, and it shows competitive results, surpassing DimeNet++ on most tasks, highlighting the importance of the DCN strategy and angle information for molecular property prediction. Note that we did not include models like EquiformerV2 in the baseline comparisons primarily because these models require significantly larger parameter counts and hardware resources; for example, EquiformerV2 has 122 million parameters, whereas our model has only 3.10 million parameters. Supporting Information Figure S2 shows the test error of DenseNGN on the IS2RES task in the OC22 challenge, demonstrating its competitive performance in direct OC22-only predictions compared to models such as SchNet, DimeNet++, PaiNN, and GemNet.

In Figure 6, we demonstrate that on the JARVIS-DFT dataset, DenseGNN outperforms the recent coGN model in most property prediction tasks. Further, in Supporting Information Figure S3, we compare DenseNGN (a nested graph network version of DenseGNN that includes angular information) on the JARVIS-DFT dataset with other nested graph network models such as coNGN, ALIGNN, and DimeNet++. DenseNGN shows superior results. These results further confirm the effectiveness of the DCN and LOPE strategies in enhancing material property prediction performance. Supporting information Figure S4 present the MAE comparison results of DenseGNN and DenseNGN with reference models on LipopDataset, FreeSolvDataset, and ESOLDataset. DenseGNN again demonstrates its competitive performance across different property prediction tasks.

In Supporting information Figure S5 showcases the performance comparison of DenseGNN-Lite against reference models on experimental small datasets. These datasets, sourced from matminer (https://github.com/dhw059/DenseGNN/datasets/dataset_metadata.json), have sample sizes ranging from 100 to 3000, covering various scales of experimental data. The results demonstrate that DenseGNN-Lite exhibits very high learning efficiency on these experimental small datasets, learning and adapting to the data more rapidly than reference models and providing superior prediction results. Figure S6 further confirms DenseGNN's learning efficiency in scenarios with small data volumes. This figure displays the learning curves for the OptB88vdW formation energy and bandgap models, with uncertainty values representing the standard error of the 5-fold cross-validation iterations. DenseGNN shows rapid learning ability, quickly converging to low prediction errors even when experimental data is scarce, further validating its learning efficiency and prediction accuracy on small datasets. Taken together, the results in Figure S5 and S6 demonstrate that DenseGNN models not only excel on computational datasets but also show efficient learning capabilities and good predictive performance on experimental small datasets. This is particularly important in materials science research, where experimental data is often limited and DenseGNN's ability to exploit limited data for more accurate predictions has significant practical implications. Figure S7 to S11 present the test results of DenseGNN, DenseGNN-Lite, and DenseNGN on the Matbench, Jarvis-DFT, and QM9 datasets.

## Ablation study

| Models | Log kvrh | GCL | % improve | Phonons | GCL | % improve |
|---|---|---|---|---|---|---|
| **DenseGNN** | **0.0512** | 5 | - | **24.8470** | 5 | - |
| Model1 | 0.0552 | 5 | -7.81 | 26.9417 | 5 | -8.43 |
| Model2 | 0.0632 | 5 | -23.44 | 32.1046 | 5 | -29.21 |
| Model3 | 0.0548 | 5 | -7.03 | 29.2872 | 5 | -17.87 |
| Model4 | 0.0571 | 5 | -11.52 | 30.4843 | 5 | -22.69 |
| DeeperDenseGNN | 0.0490 | 15 | 4.30 | 23.1206 | 15 | 6.95 |

Table1, the MAE results of ablation experiments of DenseGNN to perceive the impact of each component. Other models differ from the default architecture of DenseGNN as follows. Model 1 – LOPE node embeddings removed; Model 2 - DCN network removed; Model 3 - all DCN-GC replaced with Schnet-GC; Model 4 - all DCN-GC replaced with GIN-GC. The reference model is highlighted in bold. GCL indicates the number of GC layers.

To gain a deeper understanding of the contributions of each component in the DenseGNN to the prediction accuracy MAE, we conducted a series of ablation experiments. We selected the DFT Voigt-Reuss-Hill average bulk modulus(log kvrh) and the peak frequency of phonon DOS(phonons) as the experimental datasets. The experimental results are summarized in Table 1. Overall, the impact of the DCN component on the prediction results exceeded that of the LOPE component, leading to decrease of 23.44% and 29.21% in log kvrh and phonons predictions, respectively, while the decrease from the LOPE component were 7.81% and 8.43%, respectively. When only replacing the GC component in DenseGNN with Schnet-GC and GIN-GC, we observed a substantial decrease in prediction accuracy. On the log kvrh and phonons datasets, the prediction accuracy decreased by 7.03%, 11.52%, and 17.87%, 22.69%, respectively. Furthermore, we attempted to increase the number of GC layers in DenseGNN, and the results indicated that for log kvrh and phonons predictions, the prediction accuracy increased by 4.30% and 6.95%, respectively. Although adding more GC layers may lead to oversmoothing problems, the combination of DCN and LOPE resulted in a substantial performance improvement in our DenseGNN. Additionally, with an increase in model depth, accuracy also improved, indicating that more GC layers can more effectively capture embedding features and utilize features of higher-order neighbors. Table S1 in the supporting information presents the results of all models in ablation experiments on the jdft2d (exfoliation energy), phonons, dielectric (refractive index), perovskites (perovskite formation energy), as well as log gvrh and log kvrh (logarithm of DFT Voigt-Reuss-Hill average shear modulus and bulk modulus) datasets.

In Supplementary Figure S12 to S15, conduct ablation experiments to confirm the role of each component in the model. Figure S12 demonstrates the effect of varying the number of hidden units on the DenseGNN model's performance on the JARVIS-DFT OptB88vdW formation energy and bandgap datasets. The results show that as the number of hidden features increases from 64 to 256, the model's parameter and training time increase, but the model's MAE improves. Figure S13 explores the impact of changing the number of Graph Convolutional Network (GCN) layers on the DenseGNN model's performance on the same datasets. The results indicate that as the number of GCN layers increases, the model's parameters and training time increase, but the model's MAE improves. Figure S14 investigates the effect of altering the number of Dense Connectivity Network (DCN) layers on the DenseGNN's performance. The results show that as the number of DCN layers increases, the model's parameters and training time increase, but the model's MAE improves. The results in Figure S15 show that introducing the DCN strategy reduces the MAE test error of Schnet, PAiNN and DimeNet++ on the QM9 dataset, validating the strategy's ability to optimize the performance of current models. In the Supplementary Figures S16 to S19 provide the original 5-fold train and test results. Through these ablation experiments, we not only confirmed the role of each component in DenseGNN but also showcased how the DCN and LOPE strategies work together to improve the model's predictive accuracy and deepen its learning capabilities.

## Our DCN and LOPE strategy improve GNN models in different fields

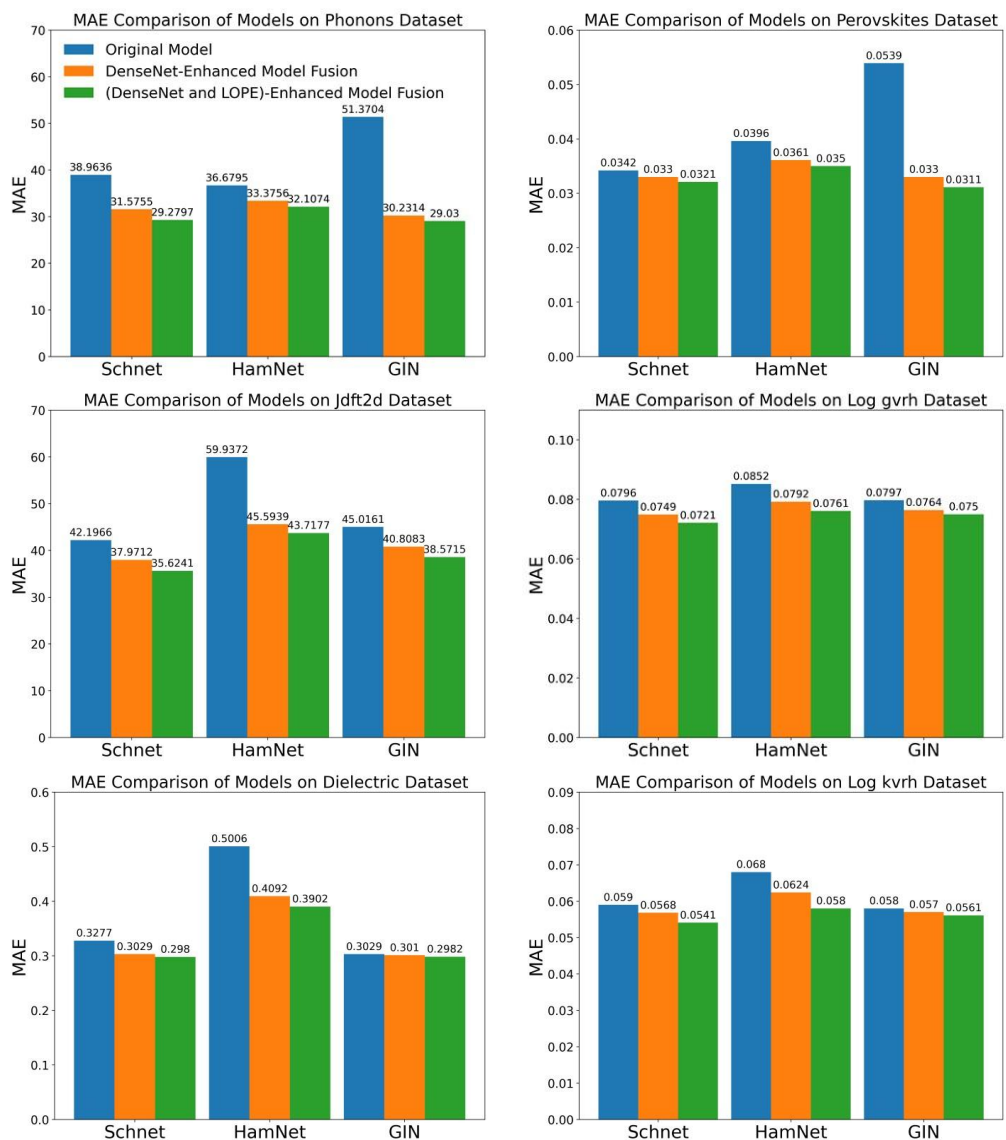

Figure 7. Comparison of the MAE changes for all GNN models on 6 different property datasets in Matbench after fusion with DCN and LOPE strategies.

Our ablation experiments highlighted the crucial roles of DCN and LOPE strategy in enhancing model performance. While numerous models have been proposed for computer, molecular, and materials fields, their performance often deteriorates when transferred to the crystal materials domain. To validate the effectiveness and generality of the DCN and LOPE strategy, we fused DCN and LOPE strategy into representative models from these three fields. We selected GraphSAGE[47], GAT[48], and GIN[49] for the computer domain; AttentiveFP[50], PAiNN[51], and HamNet[52] for the molecular domain; and CGCNN, MEGNet, and Schnet for the materials domain. These models cover spatial convolutions, message passing, 3D geometric message passing, attention mechanisms, and graph transformers, among others, in the realm of GNNs.

As shown in Figure 7, we compared the MAE results of Schnet, HamNet, and GIN models in their original state and after fusing DCN and LOPE strategy on six Matbench regression datasets (ranging from 312 to 18,928 samples). The experimental results demonstrate performance improvements across all datasets for all models. Particularly, the enhancement of model performance by DCN exceeded that of LOPE, aligning with the findings of the ablation experiments, notably prominent in models from the molecular and computer fields such as HamNet and GIN. This further confirms the impact of DCN and LOPE strategy in enhancing model cross-domain performance, demonstrating their versatility and scalability. Supplementary Figure S20 presents the MAE results comparison of representative models from the three fields fused with DCN strategy on the six Matbench datasets.

In the Matbench test results, notably on small datasets such as Jdft2d, Phonons, and Dielectric, the SchNet fused with DCN and LOPE achieved decrease in MAE. Similarly, the HamNet model exhibited notable performance improvements in Phonons and Dielectric properties, with MAE errors decreased by 23.93% and 18.26%, respectively. For the GIN model, the most performance improvements were observed on the Phonons and Perovskites datasets, with MAE errors reduced by 41.15% and 38.78%, while the decrease in MAE for other properties ranged from 3% to 10%. These results underscore the effectiveness of DCN and the LOPE strategy in enhancing the performance of models transferred from other fields to the materials domain, offering a new technical solution to address the challenge of extending existing GNNs to broader application fields.

## Fusion of DCN to address the over-smoothing problems of GNNs

| Models | GCL | Jdft2d | Phonons | Dielectric | Perovskites | Log gvrh | Log kvrh |
|---|---|---|---|---|---|---|---|
| DenseGNN | 5 | 32.5425 | 24.8470 | **0.2837** | 0.0268 | 0.0668 | 0.0512 |
| | 10 | **31.0709** | 23.6663 | 0.2980 | 0.0260 | 0.0659 | 0.0508 |
| DeeperDenseGNN | 15 | 31.8750 | **23.1206** | 0.2899 | **0.0257** | 0.0655 | **0.0490** |
| | 20 | 32.3981 | 23.3680 | 0.2993 | 0.0259 | **0.0641** | 0.0507 |
| %improve | | 4.52 | 6.95 | - | 4.10 | 4.04 | 4.30 |
| | 30 | 32.0705 | 23.8670 | 0.2752 | 0.0254 | 0.0642 | 0.0502 |
| Schnet | 5 | 35.6241 | 29.2797 | **0.2980** | 0.0321 | 0.0721 | 0.0541 |
| | 10 | 34.9571 | 28.1653 | 0.3012 | 0.0310 | 0.0710 | 0.0531 |
| DeeperSchnet | 15 | 34.6213 | 27.9956 | 0.3028 | 0.0305 | 0.0707 | **0.0521** |
| | 20 | **34.5140** | **27.9526** | 0.3112 | **0.0294** | 0.0699 | 0.0529 |
| %improve | | 3.12 | 4.53 | - | 8.41 | 3.05 | 3.70 |
| | 30 | 34.6395 | 27.8337 | 0.3036 | 0.0299 | 0.0701 | 0.0518 |
| HamNet | 5 | 43.7177 | 32.1074 | 0.3902 | **0.0350** | 0.0761 | 0.0580 |
| | 10 | 41.6590 | 30.4543 | 0.3869 | 0.0354 | **0.0730** | 0.0568 |
| DeeperHamNet | 15 | 41.3699 | **30.0089** | 0.3822 | 0.0357 | 0.0740 | 0.0565 |
| | 20 | **40.6452** | 30.0559 | **0.3781** | 0.0354 | 0.0739 | **0.0554** |
| %improve | | 7.03 | 6.54 | 3.10 | - | 4.07 | 4.64 |
| | 30 | 40.9385 | 30.1382 | 0.3748 | 0.0356 | 0.0733 | 0.0557 |
| GIN | 5 | 38.5715 | 29.030 | 0.2982 | 0.0311 | 0.0750 | 0.0561 |

|  | 10 | 36.2933 | 27.2631 | 0.2893 | 0.0302 | 0.0711 | 0.0536 |
| --- | --- | --- | --- | --- | --- | --- | --- |
| DeeperGIN | 15 | 35.7153 | 27.7466 | **0.2872** | 0.0294 | 0.0698 | 0.0527 |
|  | 20 | **35.1504** | **26.8049** | 0.2903 | **0.0291** | **0.0694** | **0.0524** |
| %improve |  | 8.87 | 7.66 | 3.69 | 6.43 | 7.46 | 6.59 |
|  | 30 | 34.3700 | 26.7400 | 0.3020 | 0.0288 | 0.0693 | 0.0520 |
|  | 40 | 35.1443 | 27.3844 | 0.2960 | 0.0286 | 0.0695 | 0.0520 |
|  | 50 | 35.8456 | 26.1842 | * | * | * | * |
|  | 60 | 35.7709 | 26.6640 | * | * | * | * |

Table 2 evaluates the impact of network depth on the MAE of DeepGNN, Schnet, HamNet, and GIN models fusing DCN and LOPE strategies across six Matbench datasets. All of these models are able to scale to at least 30 GC layers. The best results are highlighted in bold. The shaded area specifically showcases the MAE results of models with over 30 GC layers. * denotes training failure due to insufficient computational resources.

In view of the prevalent oversmoothing problem in existing message-passing GNNs, we explored whether the DCN and LOPE strategies could address this problem of and enable them to benefit from deeper GC layers. In previous studies on DeeperGATGNN[53], it was found that existing GNN models such as SchNet, CGCNN, MEGNet, and GATGNN experience a significant performance decline after adding a certain number of GC layers, leading to inaccurate property predictions. The DGN and skip connections strategies proposed in the paper did not effectively improve the scalability of models other than DeeperGATGNN, almost all models experienced a substantial performance decrease after 20 GC layers. Based on this, we implemented the DCN and LOPE strategies to the Schnet, HamNet, and GIN models across different fields and studied the scalability of these models and their deep versions (DeeperDenseGNN, DeeperSchnet, DeeperHamNet, and DeeperGIN). We conducted this study using six Matbench datasets, with all experiments employing 300 epochs and 5-fold cross-validation, training each model with 5, 10, 15, and 20 GC layers, and evaluating their scalability.

The results in Table 2 show that DeeperSchNet exhibited a performance improvement of approximately 3% to 9% across the six datasets as the number of GC layers increased from 10 to 20, indicating that our strategies effectively improved the scalability of the model. For instance, on the Perovskites dataset, the MAE decreased from 0.0321 with 20 GC layers to 0.0294 with 5 GC layers, and there was no decrease in MAE with 30 GC layers. DeeperDenseGNN, DeeperHamNet, and DeeperGIN also showed similar results, with these models demonstrating an improvement in MAE across all test datasets as the number of GC layers increased. DeeperDenseGNN improved by approximately 4% to 7%, DeeperHamNet by approximately 3% to 8%, and DeeperGIN by approximately 5% to 9%. Particularly, the DCN and LOPE strategies performed better on the GIN, with the MAE decreasing even with over 20 layers. Further testing of DeeperGIN with 30, 40, 50, and 60 GC layers revealed a continued decrease in MAE, especially notable on smaller datasets such as Jdft2d and Phonons (with sample sizes of 636 and 1265, respectively), showing improvements from 8.87% and 7.66% to 10.89% and 9.80%, respectively.

Our DCN and LOPE strategies improved the scalability of models in various fields. All these models showed a decrease in MAE across all datasets at 20 GC layers. Furthermore, even beyond 30 GC layers (see shaded area in Table 2 and Supplementary Figure S21), no model exhibited a performance decline, indicating the potential for further increasing the number of GC layers. Due to computational resource constraints, we did not continue to increase the number of layers. A key point is that our models could scale to over 60 GC layers, with performance still slightly improving, especially on smaller datasets. This suggests that with more training samples for deeper training, we have the potential to achieve better results. In summary, our experiments with 60 GC layers demonstrate that DCN and LOPE strategies can improve the scalability of models in multiple fields, with model performance not deteriorating with an increase in GC layers, demonstrating strong robustness against overfitting.

**Optimizing edges connectivity for efficiency improvement**

| \multicolumn{9}{c}{Jarvis dft_3d_formation_energy_peratom} |
|---|---|---|---|---|---|---|---|---|
| Models | ALIGNN (KNN=12) | MEGNet (r=5.0) | Schnet (r=6.0) | CGCNN (r=5.0) | DimeNetPP (r=5.0, 20) | coGN-asymmetric (KNN=24) | coNGN-asymmetric (KNN=24) | Dense GNN (KNN=6,12) | DenseGNN-Lite (KNN=6,12) |
| Graph Parameters | 7191704/124/553951/10 | 19308382/346/553941/10 | 33517592/601/553941/10 | 19308382/346/553941/10 | 11070762/199/553951/10 | 7122273/127/276931/5 | 7122273/127/276931/5 | 7191704/124/553951/10 | 7191704/124/553951/10 |
| Total parameters | 15400000 | 155073 | 316225 | 369984 | 1885958 | 697601 | 2799201 | 1717889 | 771456 |
| Trainable params | 15400000 | 155073 | 316225 | 366912 | 1885958 | 697601 | 2799201 | 1717889 | 771456 |
| MAE | 0.0331± 0.0002 | 0.0427± 0.0029 | 0.0368± 0.0021 | 0.0551± 0.0025 | 0.0528± 0.0002 | 0.0271± 0.0004 | 0.0291± 0.0005 | 0.0269± 0.0002 | 0.0276± 0.0004 |
| Training Time/epoch(s) | 110.801 | 4.839 | 4.214 | 5.798 | 11.462 | 3.012 | 13.574 | 4.447 | 3.112 |
| Inference time/epoch(s) | 111.625 | 5.175 | 4.898 | 6.777 | 12.956 | 3.421 | 14.899 | 5.289 | 3.512 |

Figure 8 compares all baseline models, DenseGNN, and DenseGNN-Lite on the Jarvis-DFT formation energy dataset. It shows the crystal graph parameters (including total edges, average edges per graph, total nodes, average nodes per graph) for each model when using their optimal edges selection method for constructing crystal graphs. Additionally, it presents the total model parameters, trainable parameters, MAE results on the test set, and the training and inference time per epoch. The red box highlights the nested graph networks, whose parameter counts exceed those of DenseGNN and DenseGNN-Lite. All tests were run on a 4090 GPU, keeping the batch size, learning rate, and other settings consistent. The average time per epoch for training and inference was calculated using the mean from 20 epochs.

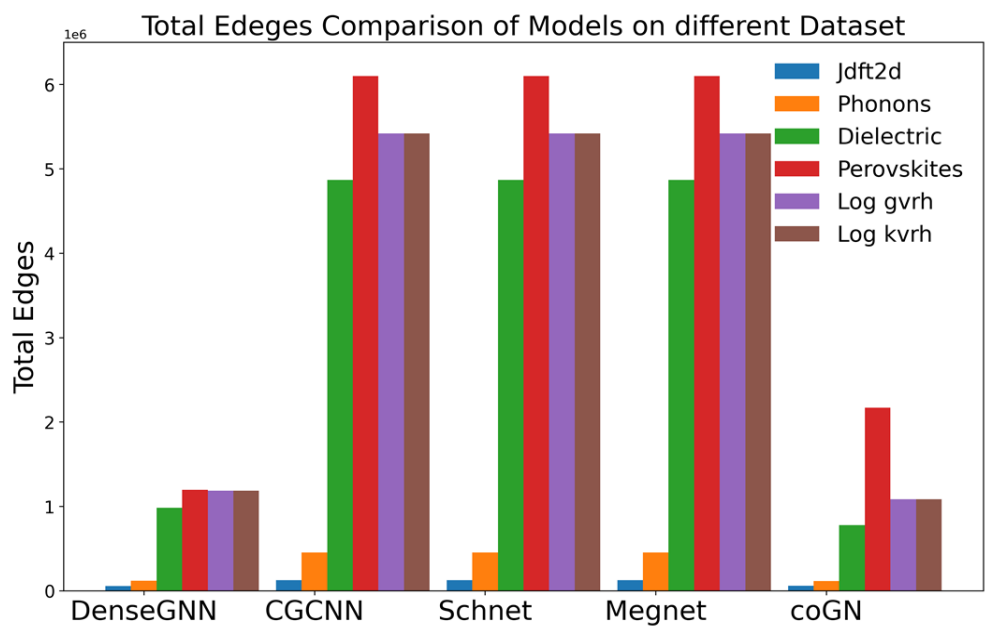

Figure 9, Comparison of edge connections among DenseGNN, SchNet, CGCNN, MEGNet, and coGN models at their optimal edges selection method with six Matbench datasets. coGN employs the method of constructing asymmetric unit graphs that fully utilize the symmetry of the crystal structure, but finds it difficult to reduce the number of nodes on the Perovskites dataset. Therefore, under the optimal MAE edges selection strategy (KNN=24), the required number of edges for coGN significantly increases.

| Perovskites | | | |
|---|---|---|---|
| Models | coGN-asymmetric (KNN=24) | DenseGNN (KNN=12) | DenseGNN-Lite (KNN=12) |
| Total graph edges | 2167618 | 1196573 | 1196573 |
| Total parameters | 697601 | 1717889 | 771456 |
| Trainable params | 697601 | 1717889 | 771456 |
| MAE | 0.0269±0.0008 | 0.0268±0.0007 | 0.0286±0.0011 |
| Training Time/epoch(s) | 1.859 | 1.005 | 0.6859 |
| Inference time/epoch(s) | 2.241 | 1.182 | 0.8840 |

Figure 10 compares coGN, DenseGNN, and DenseGNN-Lite on the Perovskites dataset from Matbench. It shows the crystal graph parameters (including total edges, average edges per graph, total nodes, average nodes per graph) for each model when using their optimal edges selection method for constructing crystal graphs. Additionally, it presents the total model parameters, trainable parameters, MAE results on the test set, and the training and inference time per epoch. The average time per epoch for training and inference was calculated using the mean from 20 epochs.

In GNNs, the nested graph networks strategy enhances the model's expressive power by incorporating angle information, but it also leads to a significant increase in the number of training parameters, thereby raising training costs. To improve efficiency, we explore more efficient methods for constructing input graphs. The selection of edge connections directly impacts model performance. As highlighted in coGN, GNNs that rely solely on relative distance information cannot distinguish geometric shapes with different angles. By strategically adding additional connections between nodes (where differences in angles can also be expressed through the distances between these additional connections, refer to Figure S24), GNNs can distinguish these shapes without relying on angle information, thus avoiding an increase in parameters. However, an excessive number of edge connections can lead to a decline in model performance.

Supplementary Figure S22 compares DenseGNN, coGN, MEGNet, SchNet, and CGCNN on 12 Matbench and Jarvis-DFT datasets, examining the changes in test set MAE as the average number of edges per graph increases using common graph construction methods such as radius-based and k-nearest-neighbors approaches. The results demonstrate that, regardless of whether the graphs are constructed using radius-based or k-nearest-neighbors methods, DenseGNN requires the fewer average edges to achieve the better MAE results. MEGNet, SchNet, and CGCNN perform optimally with the radius-based method, while coGN requires more edges than DenseGNN to achieve better MAE test results. Figure 8 compares all reference models and the DenseGNN on the Jarvis-DFT formation energy dataset, focusing on the graph parameters (including total edges, average edges per graph, total nodes, and average nodes per graph) when each model is at its optimal edges selection method for constructing crystal graphs. It also compares the models' total parameters, trainable parameters, MAE on the test set, and training and inference times per epoch. From this figure, we draw the following conclusions: First, the model parameters and the number of edges required for optimal performance by DenseGNN are lower compared to nested graph networks like ALIGNN, DimeNetPP, and coNGN. Therefore, under consistent training and testing hyperparameters (such as batch size, learning rate, and optimizer parameters), DenseGNN has the shortest training and inference times. Second, for non-nested graph networks such as SchNet, CGCNN, MEGNet, and coGN, we optimized the information passing and updating strategies for edges-nodes-graphs in DenseGNN while keeping the DCN and LOPE strategies intact. This led to the development of DenseGNN-Lite, which significantly reduced the number of trainable parameters, as shown in Figure S1. At its optimal performance, DenseGNN-Lite uses the fewer number of edges, with an average of only 124 edges per graph. Consequently, under consistent training and testing hyperparameters, DenseGNN-Lite has much shorter training and inference times.

Thanks to the DCN and LOPE strategies, DenseGNN achieves competitive performance with a lower number of required edges, thereby increasing model efficiency. The coGN strategically increases the number of edge connections using the KNN method to improve the distinguishability of the graph, and uses crystal symmetry to reduce the number of nodes in the graph to balance the number of edge connections. However, the reliance on symmetry to reduce the number of edges in the crystal/molecular graph may fail in real-world materials research, where disordered materials such as molecules and polymers are frequently encountered. As shown in Figure 9, across six Matbench properties, DenseGNN exhibits a substantially reduced number of edge connections compared to other reference models. For the perovskites, coGN uses all symmetries in constructing asymmetric unit graphs but does not reduce the average number of nodes (refer to Figure S23). Furthermore, coGN requires a higher number of edge connections (optimal k parameter of 24) to achieve optimal model performance and enhance graph discriminability (refer to Figure S24). Therefore, under the optimal MAE edges selection strategy, the required number of edges for coGN significantly increases. Figure 10 compares the graph parameters, total parameters, trainable parameters, and MAE results on the test set for coGN, DenseGNN, and DenseGNN-Lite models on the Perovskites dataset, when each model is at its optimal edges selection method for constructing crystal graphs. It reveals that, on the perovskites dataset, DenseGNN and DenseGNN-Lite have the shortest training and inference times per epoch. Supplementary Information Table S2 provides a comparison of all edges and nodes for DenseGNN, SchNet, CGCNN, MEGNet, and coGN on the Matbench dataset.

**Crystal structure distinguishment improvement**

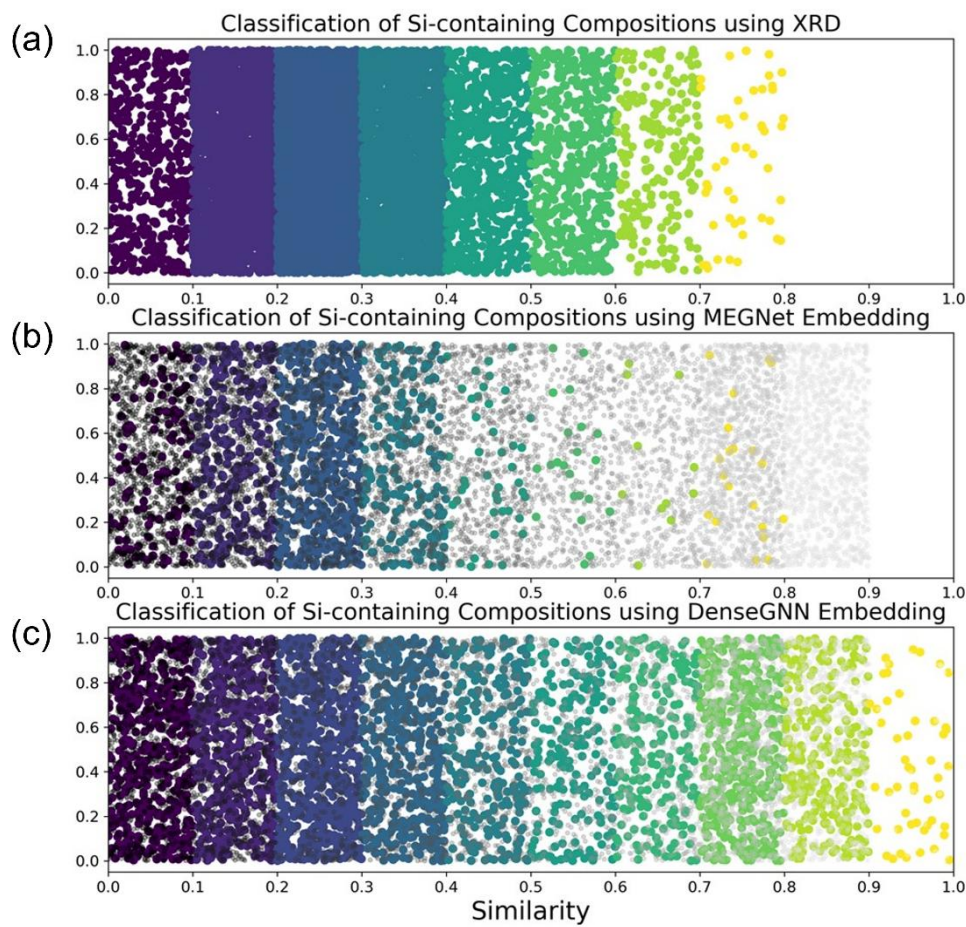

Figure 11. performance comparison in structural distinguishment of silicon-containing compounds. (a) Classification results using standard XRD method;(b) Classification results from pre-trained MEGNet;(c) Classification results from pre-trained DenseGNN. Gray dots: Misclassified structures compared to XRD results

Our approach involves extracting structural embeddings from specific layers of pre-trained models and utilizing a similarity calculation strategy similar to the standard XRD method to distinguish among over 8970 silicon-containing compounds from the Materials Project(MP) database[54]. As depicted in Figure11, our DenseGNN has shown significant improvement in distinguishing structures compared to the MEGNet. The horizontal coordinates of each colored point in the figure represent the similarity between structures, with a resolution of 0.1 and divided into 10 categories ranging from 0 to 1, where structures become increasingly similar from left to right. Figure 11(a) depicts the classification results of silicon-containing structures using the standard XRD method, while Figure 11(b) presents the classification results based on the structural embeddings derived from the MEGNet pre-trained on the bandgap dataset from the MP database[23]. Figure 11(c) showcases the classification results using the structural embeddings obtained from our DenseGNN pre-trained on the same dataset. The gray points in the figure represent structures where the pre-trained models' classifications are inconsistent with the standard XRD results.

It is evident that the DenseGNN has shown a significant improvement in classification accuracy compared to the MEGNet. The classification accuracy of MEGNet is less than 40% (with the majority of errors occurring on the right side of the figure, involving highly similar structures, which aligns with common knowledge that very similar structures are indeed challenging to distinguish), while the accuracy of DenseGNN has nearly doubled, reaching approximately 80%. This results highlights the superior performance of our model in distinguishing material structures.

**Discussion**

In this study, we propose the DenseGNN to address a series of problems in material property prediction faced by existing GNNs. Despite the introduction of various GNNs such as CGCNN, MEGNet, ALIGNN, and coGN since 2018, which have made substantial achievements in aspects like global state variable representation, incorporation of bond angle information, and optimization of crystal graph edge connections, there are still some problems that need to be overcome. First, we observe that nested graph networks like ALIGNN and coGN, while effective in certain cases, are limited in their widespread application due to high training costs and advantages on only specific datasets. To tackle these problems, we introduce LOPE to optimize the atomic embedding representation and edge connectivity, effectively learn multi-body interactions such as bond angles and local geometric distortions, and achieve minimal edge connections. This reduces the training time required for larger GNNs without sacrificing accuracy. Second, we introduce the DenseGNN model to address the problems of extending GNN to diverse applications in materials, molecular, and chemical fields. We specifically focus on the generalization ability and performance trade-offs of existing GNN models across different fields. While GNNs perform well on specific domain datasets, they often struggle to maintain consistent performance when applied across different fields, limiting their potential in broader fields. DenseGNN is fused with DCN, HRN, and the LOPE strategy, leading to performance improvements across multiple fields. DenseGNN not only achieves optimal performance on benchmark datasets in molecular, crystal materials, and catalysis fields, demonstrating its versatility and scalability across different fields, but also outperforms existing coGN and MODNET on small datasets, showcasing its exceptional generalization capabilities across various datasets. Third, we apply our DCN and LOPE strategies to GNN in fields such as computer science, crystalline materials, and molecules, resulting in performance improvements on the Matbench dataset. Finally, we address the problems of over-smoothing in GNNs, which is a major barrier to increasing the number of model layers. The DenseGNN overcomes this problem by fusing the DCN design, enabling the construction of very deep networks while avoiding performance degradation.

The DenseGNN has achieved success in the fields of materials science, molecular, and chemistry, demonstrating its advantages in handling complex many-body interactions and improving training efficiency. Despite the outstanding performance of DenseGNN on benchmark datasets in multiple domains, we recognize that there are still some potential problems and challenges that need to be addressed in future research. First, the deep network structure of DenseGNN may face higher computational complexity when dealing with large-scale graph data. To overcome this challenge, future research can explore more efficient structural graph optimization strategies and model architecture designs to reduce the computational cost. Additionally, the "black box" nature of DL models also limits the interpretability of the model. Therefore, developing new visualization tools and techniques to aid in understanding the decision-making process of DenseGNN will be crucial in improving the transparency of the model. Although DenseGNN is versatile, further optimization or integration of domain-specific pre-training knowledge may be necessary in specific domains. Developing customized DenseGNN tailored to the data characteristics and application requirements of specific domains will help improve the model's performance on specific tasks. In conclusion, the introduction of the DenseGNN model provides a new perspective and tool for the application of GNN in multiple domains. By addressing these problems mentioned above, we look forward to DenseGNN achieving broader applications in future research and driving scientific advancements in related fields.

## Methods

### Data description

The JARVIS-DFT dataset was developed using the Vienna ab initio simulation package (VASP). Most properties are calculated using the OptB88vdW[55] functional. For a subset of the data, we use TBmBJ[56] to get a better bandgap. We use density functional perturbation theory (DFPT)[57] to predict piezoelectric and dielectric constants with electronic and ionic contributions. The linear response theory-based[58] frequency based dielectric function was calculated using OptB88vdW and TBmBJ, and the zero energy values are used to train ML models. The TBmBJ frequency-dependent dielectric function is used to calculate the maximum efficiency limited by the spectrum (SLME)[59]. The magnetic moment is calculated using spin-polarized calculations considering only ferromagnetic initial configurations and ignoring any density functional theory (DFT) + U effects. Thermoelectric coefficients such as the Seebeck coefficient and power factor are calculated using the BoltzTrap software[60] with a constant relaxation time approximation. The exfoliation energy of van der Waals bonded two-dimensional materials is calculated by calculating the energy difference between each atom in the bulk phase and the corresponding monolayer. Spin orbit spillage[61] is computed as the disparity between material wavefunctions with and without the inclusion of spin-orbit coupling effects. All JARVIS-DFT data and classical force field inspired descriptors (CFID)[62] are generated using the JARVIS-Tools software package. The CFID baseline model is trained using the LightGBM software package[62].

Matbench is an automated benchmark testing platform specifically designed for the field of materials science, designed to evaluate and compare the most advanced ML algorithms that predict various solid material properties. It provides 13 carefully curated ML tasks that cover a wide range of inorganic materials science, including the prediction of various material properties such as electronics, thermodynamics, mechanics to thermal properties of crystals, two-dimensional materials, disordered metals, etc. The datasets for these tasks come from different density functional theories and experimental data, with sample sizes ranging from 312 to 132,000. The platform is hosted and maintained by the MP, providing a standardized evaluation benchmark for the field of materials science.

QM9 provides molecular properties calculated by DFT, such as highest occupied molecular orbital (HOMO), lowest unoccupied molecular orbital (LUMO), energy gap, zero-point vibrational energy (ZPVE), dipole moment, isotropic polarizability, electron spatial extent, internal energy at 0K, internal energy at 298K, enthalpy at 298K, Gibbs free energy at 298K, and heat capacity. LipopDataset: Lipophilicity is an important feature that affects the membrane permeability and solubility of drug molecules. This lipophilicity dataset is curated from the ChEMBL database, providing experimental results of octanol/water partition coefficients (logD at pH 7.4) for 4200 compounds. FreeSolvDataset: The FreeSolv dataset consists of experimental and computationally derived solvation free energies of small molecules in water, along with their experimental values. Here, we utilize a modified version of the dataset that includes the SMILES strings of the molecules and their corresponding experimental solvation free energy values. ESOLDataset: The Delaney (ESOL) dataset is a regression dataset containing structures and water solubility data of 1128 compounds. This dataset is widely used to validate the ability of ML models to directly estimate solubility from molecular structures encoded as SMILES strings.

The OC22 dataset focuses on oxide electrocatalysis. A crucial difference between OC22 and OC20 is that the energies in OC22 are DFT total energies. DFT total energies are more challenging to predict but offer the most generality and are closest to a DFT surrogate, providing flexibility to study property prediction beyond adsorption energies. Similar to OC20, the tasks in OC22 include S2EF-Total and IS2RE-Total.

## DenseGNN implementation and training

### DCN

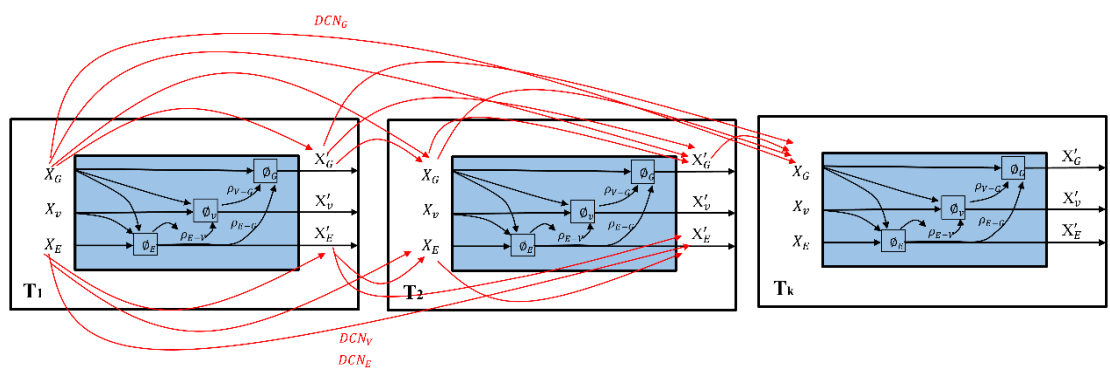

The process of densely connecting all GC layers based on the DCN strategy in DenseGNN is illustrated in Figure 12. In each graph convolutional layer (GC layer), the edge-node-graph is densely connected to the corresponding edge-node-graph in all preceding GC layers. This design allows each GC layer to fully utilize the feature information from the preceding GC layers, enabling effective feature reuse. Additionally, by concatenating the feature maps of all layers, the shortest information propagation path is achieved. All dense connection channels are highlighted in red.

We described the process of densely connecting all GC layers based on the DCN strategy as shown in Figure 12. In comparison to ResNet, DCN proposes a more aggressive dense connection mechanism: that is, each layer will connect to all preceding layers, specifically, each layer will take all preceding layers as additional inputs. For example, the input of $x_E{'}$ in layer T$_2$ includes not only $x_E$ from layer T$_2$, but also $x_E$ from preceding layer T$_1$ and $x_E{'}$, all concatenated along the last dimension. Unlike ResNet, where each layer is connected to a preceding layer through element-wise addition shortcut connections, in DCN, each GC layer concatenates with all preceding layers at the edge-node-graph level and serves as input to the next layer. For a network with k GC layers, DCN includes a total of *k(k+1)/2* connections, which is a dense connection compared to ResNet. Additionally, DCN directly connects feature maps from different layers, enabling feature reuse and improving efficiency, which is the main difference between DCN and ResNet. In the formula representation, the output of a traditional network at the *k*-th layer is:

$$x_k = H_k(x_{k-1})$$

For ResNet, an identity function from the preceding layer input is added:

$$x_k = H_k(x_{k-1}) + x_{k-1}$$

In DCN, all preceding layers are concatenated as input:

$$x_k = H_k([x_0, x_1, \ldots, x_{k-1}])$$

Here, $H_k(.)$ represents the non-linear transformation function in the GC layer, which is a composite operation that include a series of connection operations, a 3-layer MLP network, residual operations, and the Swish activation function, applied to the edges, nodes, and graph objects, achieving synchronous updates of edge-node-graph features.

In this study, we innovatively integrate the design philosophy of DCN with GNN to enhance the performance of GNN in material property prediction tasks. The core of this fusion strategy lies in the dense connectivity feature of DCN, which brings advantages to the GNN model. First, the dense connectivity strategy of DCN ensures that each layer of GC in DenseGNN can directly access the feature information of all preceding GC layers. This design allows each GC layer to fully utilize the information of edges, nodes, and graph levels, achieving efficient information propagation, reducing the risk of information loss, enhancing network training efficiency, and achieving more precise feature representation. Second, DCN concatenates edge, node, and graph-level features in the channel dimension, enabling direct information propagation between GC layers. This design not only simplifies the network structure but also accelerates the flow of information between GC layers. This direct information flow mechanism helps improve the GNN's ability to capture complex graph structural features, especially in dealing with large-scale graph data, enabling more effective feature extraction and pattern recognition. Last, DCN's feature reuse mechanism improve the learning ability of the GNN model. This design allows the network to access and utilize feature information from all preceding GC layers at each layer, thereby improving the model's ability to capture data patterns. Additionally, as feature reuse reduces the need for additional parameters, it helps reduce the model's complexity and the risk of overfitting.

In summary, the fusion of DCN with GNN not only improves the model's performance but also improve the robustness and generalization ability of the model in handling complex graph data. This fusion strategy provides a new perspective for research in the field of materials science, with the potential to drive the discovery of new materials.

**LOPE**

We employed LOPE to represent node features, optimizing the atomic embedding representation, and enhancing the performance of DenseGNN in material property prediction.

LOPE includes atomic embeddings and orientation-resolved embeddings, calculated through the product integration of Radial Distribution Functions (RDFs) and Gaussian window functions to describe the local atomic environment. The computation process of LOPE is as follows:

1. Initialization of parameters: we set the directions, Gaussian window widths, and cutoff distances. The direction parameter can be None or specific coordinate axes (such as 'x', 'y', 'z'), while the Gaussian window width is typically a list of floating-point numbers defining different window widths. The cutoff distance is a threshold used to determine the interaction range between atoms.

2. Neighbor atoms and distance calculation: For a given atom in the structure, we calculate all its neighboring atoms and determine the distances between them.

3. Direction-dependent processing: If direction-dependent fingerprints need to be computed, we calculate the displacement of each neighboring atom relative to the central atom.

4. Calculation of the cutoff function: We use the cutoff function f(r) to limit the interactions between atoms, where f(r) is $0.5[\cos(\frac{\pi r}{R_c}) + 1]$ when the distance r is less than the cutoff distance Rc, and 0 when r is greater than or equal to Rc.

5. Calculation of Gaussian windows: For each Gaussian window width value, we compute the Gaussian window function, which is the product of the exponential function of the distance squared divided by the Gaussian window width, multiplied by the cutoff function Rc.

6. Calculation of fingerprints: Based on the direction parameter, we compute the atomic fingerprints. For non-directional fingerprints, we sum all window function values. For directional fingerprints, we multiply the window function by the corresponding component of displacement, and then sum them.

7. Output results: Finally, we horizontally stack all computed fingerprint vectors to form the final feature vector.

The introduction of the LOPE strategy improved the efficiency of the DenseGNN in material data learning. This strategy optimizes the atomic embedding representation, allowing the model to achieve efficient training with minimal edge connections while maintaining prediction accuracy. Furthermore, benefiting from the effectiveness of LOPE, the DenseGNN can construct deeper networks with up to 60 layers, and as the network depth increases, the model performance shows a steady improvement trend.

**Implementation details**

In this study, we utilized the TensorFlow and Keras deep learning frameworks to construct all models. During the implementation process, we relied on a series of important libraries, including KGCNN, Python Materials Genomics (Pymatgen)[63], the RDKit open-source cheminformatics toolkit[64], and PyXtal[65]. The training of all models was trained on an NVIDIA RTX 4090 24GB GPU. Regarding the model parameter settings, the default cutoff radius for Schnet, CGCNN, and MEGNet was set to 6 angstroms. The default cutoff radius for HamNet and GIN was set to 5 angstroms, while the maximum number of neighbors for nodes (excluding self, as self-loops are allowed) was limited to 17. When dealing with edges in the graph, we first computed the distance matrix between nodes, then selected edges based on the cutoff radius as a threshold, limiting the number of neighbors to within the default value of 17. For DenseGNN, we employed the KNN edge selection method, with the parameter k set to 12 for optimal performance. The KNN method relies on the number of neighbors k, but edge distances may vary with changes in crystal density. Subsequently, we applied a Gaussian kernel function to extend edge lengths and used them as the edge features of our model's graph. For the node features of DenseGNN, we introduced the LOPE representation as a 24-dimensional vector added to the node features. To evaluate model performance, we used the MAE as the standard evaluation metric, which is the common evaluation method for material property prediction and the primary evaluation metric used in this study for all models and Matbench benchmark tests. We assessed the performance of all models based on specific experimental designs using 5-fold cross-validation and hold-out testing methods.

# Data availability

All data including matbench, jarvis-DFT, QM9, OC22, and experimental datasets used in this work are available at the Github link https://github.com/dhw059/DenseGNN/blob/main/datasets/.

# Code availability

The code and training configurations for different versions of DenseGNN and comparative models, as well as the result plotting scripts, are available on GitHub at https://github.com/dhw059/DenseGNN/.

# Acknowledgements

We are grateful for the financial support from the National Key Research and Development Program of China (Grant Nos.2021YFB3702104). The computations in this paper were run on the π 2.0 cluster supported by the Center for High Performance Computing at Shanghai Jiao Tong University.

# Author information

**Authors and Affiliations**
School of Materials Science and Engineering, Shanghai Jiao Tong University, Shanghai 200240, China.
Zhangjiang Institute for Advanced Study, Shanghai Jiao Tong University, Shanghai 201203, China.
Materials Genome Initiative Center, Shanghai Jiao Tong University, Shanghai 200240, China.
Hongwei Du, Jian Hui , Lanting Zhang, Hong Wang

School of Materials Science and Engineering, Shanghai Jiao Tong University, Shanghai 200240, China.
Jiamin Wang

**Contributions**

H.D and H.W devised the idea for the paper. H.D implemented the idea and conducted the code design and visualizations. H.D and H.W interpreted the results and prepared the manuscript. J.H, L.Z, and J.W contributed to the data analysis and provided critical feedback on the manuscript.

**Corresponding author**

Correspondence to Hong Wang.

# Ethics declarations

Competing interests

The authors declare no competing interests.

# Supplementary information

Overview of the DenseGNN and DenseGNN-Lite architecture; Test MAE for multiple tasks of the OC22 and JARVIS-DFT datasets on DenseNGN and reference models; Test MAE for multiple tasks of the Lipop, FreeSolv and ESOL benchmark datasets on DenseNGN and reference models; Test MAE for DenseGNN-Lite and reference models on experimental small datasets; Learning curve for OptB88vdW formation energy and bandgap models; Figures S7 to S11 present the test results of DenseGNN, DenseGNN-Lite, and DenseNGN on the Matbench, Jarvis-DFT, and QM9 datasets; Table S1 in the supporting information presents the results of all models in ablation experiments on the Matbench datasets. In Supplementary Figure S12 to S15, conduct ablation experiments to confirm the role of each component in the model. Figures S16 to S19 provide the original 5-fold train and test results; Figure S20: MAE comparison of representative models in three domains (materials, molecular, and computer science) on 6 Matbench datasets before and after fusing the DenseNet strategy; Figure S21 presents the learning curves of GIN with 40 GC layers; Table S2 presents the comparison of all edges and nodes for DenseGNN, Schnet, CGCNN, MEGNet, and coGN on the Matbench dataset; Figures S22 to S24 compare MAE performance across models and graph construction methods, highlight DenseGNN's efficiency, and illustrate the importance of angle information in distinguishing geometric graphs.